\documentclass[aps,pre,superscriptaddress,showpacs,twocolumn,longbibliography]{revtex4-1}
\usepackage{color}
\usepackage[usenames,dvipsnames]{xcolor}
\usepackage{graphicx}
\usepackage{color}
\usepackage{epsfig}
\usepackage{bm} 
\usepackage{amsmath}
\usepackage{amsfonts}
\usepackage{amssymb}
\usepackage{mathtools}

\usepackage{color}
\definecolor{myblue}{RGB}{65,105,225}
\definecolor{mygreen}{RGB}{34,139,34}
\definecolor{myorange}{RGB}{255,69,0}
\usepackage[colorlinks,linkcolor=myorange,urlcolor=myblue,citecolor=myblue]{hyperref}

\usepackage[usenames,dvipsnames]{xcolor}
\usepackage{color}

\def\(({\left(}
\def\)){\right)}
\def\[[{\left[}
\def\]]{\right]}

\newcommand{\cL}{{\mathcal L}}

\newcommand{\be}{\begin{equation}}
\newcommand{\ee}{\end{equation}}
\newcommand{\beq}{\begin{equation}}
\newcommand{\eeq}{\end{equation}}
\newcommand{\ben}{\begin{eqnarray}}
\newcommand{\een}{\end{eqnarray}}

\newcommand{\vrr}{{\bf{r}}}

\newcommand{\vnabla}{{\bm{\nabla}}}
\newcommand{\vj}{{\bf{j}}}

\newcommand{\vv}{{\bf{v}}}
\newcommand{\vJJ}{{\bf{J}}}
\newcommand{\vlamb}{{\bm{\lambda}}}

\newcommand{\vxperp}{{\bm{x_{\perp}}}}
\newcommand{\vJperp}{{\bm{J_\perp}}}

\newcommand{\TJ}{{T_\vJJ}}
\newcommand{\vjJ}{{\vj_\vJJ}}
\newcommand{\vlJ}{{\bm{\lambda}_\vJJ}}
\newcommand{\psiJ}{{\psi_\vJJ}}
\newcommand{\cn}{{\text{cn}}}
\newcommand{\sn}{{\text{sn}}}
\newcommand{\dn}{{\text{dn}}}
\newcommand{\am}{{\text{am}}}

\begin{document}

\title{Infinite family of universal profiles for heat current statistics in Fourier's law}

\author{P.L. Garrido}
\email[]{garrido@onsager.ugr.es}
\affiliation{Departamento de Electromagnetismo y F\'{\i}sica de la Materia, and Instituto Carlos I de F{\'\i}sica Te{\'o}rica y Computacional. Universidad de Granada. E-18071 Granada. Spain }

\author{P.I. Hurtado}
\email[]{phurtado@onsager.ugr.es}
\affiliation{Departamento de Electromagnetismo y F\'{\i}sica de la Materia, and Instituto Carlos I de F{\'\i}sica Te{\'o}rica y Computacional. Universidad de Granada. E-18071 Granada. Spain }

\author{N. Tiz{\'o}n-Escamilla}
\email[]{tizon@onsager.ugr.es}
\affiliation{Departamento de Electromagnetismo y F\'{\i}sica de la Materia, and Instituto Carlos I de F{\'\i}sica Te{\'o}rica y Computacional. Universidad de Granada. E-18071 Granada. Spain }

\date{\today}

\begin{abstract}
Using tools from large deviation theory, we study fluctuations of the heat current in a model of $d$-dimensional incompressible fluid driven out of equilibrium by a temperature gradient. We find that the most probable temperature fields sustaining atypical values of the global current can be naturally classified in an infinite set of curves, allowing us to exhaustively analyze their topological properties and to define universal profiles onto which all optimal fields collapse. We also compute the statistics of empirical heat current, where we find remarkable logarithmic tails for large current fluctuations orthogonal to the thermal gradient. Finally, we determine explicitly a number of cumulants of the current distribution, finding interesting relations between them.
\end{abstract}
\maketitle

\section{Introduction}

The development of a theory of fluctuations in fluids has been a central object of study in statistical physics \cite{boon91a,de-groot13a}. A general framework to characterize fluctuations in thermodynamic equilibrium states was provided by Landau and Lifshitz \cite{landau13a, landau13b}, and this program has been generalized with success to study small fluctuations for fluids in nonequilibrium steady states \cite{zarate06a}. Nevertheless, understanding \emph{arbitrary} fluctuations in fluids far from equilibrium still remains an open problem, and this is the focus of the present work. An interesting situation to analyze in this context is the problem of heat transport in a fluid subject to a thermal gradient, possibly one of the \emph{simplest} and most studied cases of a nonequilibrium steady state \cite{bonetto00a}. Heat transport in this setting is governed by Fourier's law, which establishes the propotionality between the heat current and the local temperature gradient. The proportionality constant defines the heat conductivity $\kappa$, an intrinsic property of the fluid which could depend on the local temperature and density. Interestingly, while it is widely believed that Fourier's law is just a linear approximation to a more complex transport law, recent works have shown that, at least for some fluid models, this law holds locally far from equilibrium \cite{hurtado16a} and well beyond the linear transport regime. Numerous experimental works have studied the statistics of fluctuations of heat flux and temperature in this setting for a wide variety of systems, measuring the corresponding probability distributions \cite{tuttle05a,hall00a,zhao09a}, some low- and high-order cumulants \cite{katsaros77a,gollub91a,hayashi89a,coppin86a,hall00a,Krishnamurthy16a,garnier09a,zhao09a,mcphee92a,vickers97a,champagne77a} and the associated temperature profiles \cite{katsaros77a,gollub91a,coppin86a,champagne77a,garnier09a,mound17a}. Nevertheless, developing a general theoretical scheme to understand both typical and rare heat current fluctuations in this setting remains challenging, even for the simplest model fluids.

In recent years a series of works have analyzed current fluctuations in a broad family of stochastic models of transport \cite{sasa08a,bertini15a,bertini01a,bertini02a,bertini05a,bertini06a,bertini12a,bertini13a,bertini02a,prados11a,prados12a,hurtado13a,jack15a,bodineau10a,krapivsky14a,tizon-escamilla17a,perez-espigares16a,derrida98b}, offering a deeper comprehension of nonequilibrium fluctuating behavior. A key tool in these developments has been the Macroscopic Fluctuation Theory (MFT) of Bertini and coworkers \cite{bertini15a,derrida07a,hurtado14a,bertini01a,bertini02a,bertini05a,bertini06a,bertini12a,bertini13a,bertini02a}, which describes dynamical fluctuations in diffusive equilibrium and nonequilibrium media starting from their fluctuating hydrodynamic description in terms of two transport coefficients, the \emph{diffusivity} and the \emph{mobility}. In particular, MFT offers (1) explicit variational formulas for the \emph{large deviation functions} (LDFs) that control the statistics of fluctuations \cite{touchette09a,derrida07a,hurtado14a,barre18a}, and (2) differential equations for the \emph{optimal trajectory} (or sequence of configurations) adopted by the system to sustain a given fluctuation. Understanding the properties of these LDFs and the optimal trajectories is a task of crucial relevance since they contain information on interesting new physics, such as the emergence of order at the fluctuating level via dynamical phase transitions \cite{hurtado14a,bodineau05a,bodineau07a,hurtado11a,perez-espigares13a,tizon-escamilla17b,vaikuntanathan14a,lam09a,chandler10a,baek17a,baek18a,perez-espigares18a,nyawo16a,karevski17a,zarfaty16a,garrahan07a,garrahan09a}, or the appearance of new symmetries and fluctuation theorems out of equilibrium \cite{evans93a,evans94a,gallavotti95a,gallavotti95b,kurchan98a,lebowitz99a,andrieux07a,hurtado11a,villavicencio14a,lacoste14a,gaspard13a,perez-espigares15a}. The MFT equations for current statistics lead to a complex variational problem in general, so most studies to date have focused on oversimplified one-dimensional ($1D$) models, where calculations are somewhat simpler. It has been only in recent years that the MFT problem for $d>1$ diffusive systems has been tackled \cite{hurtado14a,hurtado11a,akkermans13a,villavicencio14a,kumar15a,becker15a,perez-espigares15a,perez-espigares16a,villavicencio16a,tizon-escamilla17a,tizon-escamilla17b}, and these works have shown that lifting the dimensionality constraint leads to a rich phenomenology not present in one dimension.

With these ideas in mind, the aim of this work is to characterize within the MFT framework the statistics of fluctuations of the empirical heat current in an incompressible quiescent $d$-dimensional model fluid subject to a boundary temperature gradient \cite{landau13a,zarate06a}. In particular our goal consists in describing the optimal temperature field sustaining a given heat flux fluctuation in the long-time limit, as well as determining some cumulants of the current distribution and its tail behavior. With this purpose, we solve the MFT problem for a model fluid characterized by a constant thermal diffusivity $D(T)=1/2$ and a quadratic mobility $\sigma(T)=T^2$, a model that captures the heat transport properties of a large family of quiescent incompressible fluids under moderate temperature gradients \cite{landau13a,landau13b,zarate06a}. In particular, we use the Weak Additivity Principle \cite{perez-espigares16a,tizon-escamilla17a} as a tool to obtain both the heat flux LDF and the optimal temperature field associated to each current fluctuation. We find that these optimal temperature fields can be gathered into families characterized by the same functional form (in terms of inverse Jacobi elliptic functions). This observation allows us to classify all optimal trajectories in an infinite set of universal functions, providing a deeper understanding of their properties and structure. Moreover, we obtain the analytical form of the current LDF and analyze its behavior in limiting cases, both near the steady state and in the far tails of the distribution. The latter case exhibits an interesting logarithmic dependence which confirms the complex analytic behavior of the heat current LDF. We further determine the cumulant generating function of the current distribution, from which analytical expressions for its cumulants follow, as well as relations between them which open the door to further experimental research on this problem.

\section{Model, fluctuating hydrodynamics and path integral representation}
\label{sectmodel}

We consider a $d$-dimensional fluid subject to a boundary temperature gradient in one direction, say $x\in[0,L]$ with $L$ the system linear size. The fluid is fully described at any instant of time by the mass density $\rho(\vrr,t)$, temperature $T(\vrr,t)$, pressure $p(\vrr,t)$ and local center-of-mass velocity $\vv(\vrr,t)$ fields, with $\vrr\in\Lambda\equiv[0,L]^d$ and $t>0$ the spatial and temporal coordinates, respectively. The fluid's evolution at the macroscale is completely characterized by a set of $d+2$ partial differential equations, called {\it balance equations}, which are derived from the local conservation laws together with the usual constitutive relations between the thermodynamic forces and the fluxes \cite{de-groot13a}. In particular, conservation of mass leads to the continuity equation
\begin{equation}
 \partial_t\rho+\vnabla\cdot(\rho \vv)=0\,\,,
 \label{masscons}
\end{equation}
while momentum conservation yields the Navier-Stokes equations
\begin{equation}
\rho\left[\partial_t \vv+(\vv\cdot\vnabla)\vv\right]=-\vnabla\cdot p+\eta\vnabla^2 \vv+\left(\zeta+\frac{1}{3}\eta\right)\vnabla(\vnabla\cdot \vv)\,\,,
\label{navierstokes}
\end{equation}
and conservation of energy results in
\begin{equation}
\partial_t\left(\frac{1}{2}\rho \vv^2+\rho e\right)=-\vnabla\cdot\left[\rho \vv\left(\frac{1}{2}\vv^2+\omega\right)+\bm{\Phi}+\vj_{\bf{D}}\right]\,\,.
\label{energycons}
\end{equation}
In the above equations $\eta$ and $\zeta$ are respectively the shear and bulk viscosity coefficients, $e$ is the internal energy per mass unit, $\omega$ is the enthalpy per mass unit, $\bm{\Phi}$ is the viscous dissipation function (proportional to the divergence of the velocity) and $\vj_{\bf{D}}$ is the local heat current \cite{landau13a,zarate06a,de-groot13a}. In particular, the structure of the local heat current field is given by the well-known Fourier's law of heat conduction
\begin{equation}
\vj_{\bf{D}}(\vrr,t)=-\kappa(T)\vnabla T(\vrr,t)\,\,,
\label{fourierlaw}
\end{equation}
with $\kappa(T)$ the thermal conductivity. In this paper we are interested in studying thermal transport in a \emph{quiescent incompressible fluid} in contact with two boundary thermostats at temperatures $T_0$ and $T_1$ along the $x$-direction, with periodic boundary conditions along all perpendicular $(d-1)$-directions. Quiescence implies that $\vv(\vrr,t)=0$ $\forall\,\vrr,t$, while incompressibility implies that the fluid's mass density and pressure fields are constant across space, so the only relevant field in this case is the temperature field $T(\vrr,t)$, which then satisfies Fourier's heat equation \cite{landau13a,landau13b,zarate06a,de-groot13a}
\begin{equation}
\partial_t T(\vrr,t)= \vnabla\cdot\left(D(T)\vnabla T(\vrr,t) \right)\,\,,
\label{f1}
\end{equation}
where $D=\frac{\kappa}{\rho c_p}$ is the \emph{thermal diffusivity}, with $c_p$ the specific heat at constant pressure. Finally, we further assume that the initial condition is such that the system relaxes to its steady state in a finite time scale.

The previous description is a macroscopic one. At a more interesting mesoscopic level, molecular-scale chaotic motions leave a fingerprint in the form of  small fluctuations of the heat current field \cite{landau13a,zarate06a}. This can be taken into account in a (fluctuating) hydrodynamic description by adding a (weak) noise term to the current which reflects all the fast microscopic degrees of freedom which are integrated out in the coarse-graining procedure leading to this irreversible evolution equation. The amplitude of this noise term is nontrivial, as it is coupled to the thermal diffusivity via a fluctuation-dissipation theorem which guarantees the correct equilibrium state in the absence of driving \cite{landau13a,zarate06a}. In this way the instantaneous fluctuating heat current field can be written as $\vj(\vrr,t)=\vj_{\bf D}(\vrr,t)+{\bm\xi}(\vrr,t)$, and the temperature field now obeys a stochastic evolution equation
\begin{equation}
\partial_t T(\vrr,t)+\vnabla\cdot\left[-D(T)\vnabla T(\vrr,t) + {\bm\xi}(\vrr,t) \right] = 0\,,
\label{f4}
\end{equation}
where ${\bm\xi}(\vrr,t)$ is a Gaussian white noise vector field with
\ben
\langle{\bm\xi}(\vrr,t)\rangle &=& 0\, , \label{f6} \\
\langle{\xi}_\alpha(\vrr,t){\xi}_\beta(\vrr',t')\rangle &=& \frac{1}{\Omega} \sigma[T(\vrr,t)]\delta_{\alpha\beta}\delta(t-t')\delta(\vrr-\vrr')\, , \nonumber
\een
and $\alpha,\beta\in[1,d]$. The amplitude $\sigma[T(\vrr,t)]$ is the {\it mobility} transport coefficient, coupled to the thermal diffusivity via a local Einstein relation, $D(T)=\sigma(T) f_0''(T)$, with $f_0(T)$ the \emph{equilibrium} free energy density of the fluid and $''$ denoting second derivative with respect to the function argument. Moreover, $\Omega \equiv \epsilon^{-d}$ is a (large) parameter controlling the strength of the noise that arises because of the law of large numbers when rescaling space and time diffusively as $\vrr\rightarrow\epsilon^{-1}\vrr$ and $t\rightarrow\epsilon^{-2}t$, respectively, in the coarse-graining from microscopic to mesoscopic scales \cite{spohn12a}. Indeed $\Omega$ can be interpreted as the volume of the microscopic region which is averaged to obtain the local field value at the mesoscale, and the limit $\Omega\to \infty$ corresponds to the macroscopic hydrodynamic description of the fluid. Here we are interested in the (weak noise) limit of large but finite $\Omega$, relevant to understand fluctuations in nanosize systems.

The usual way to proceed now in order to study the properties of a fluid's fluctuations would consist in linearizing the stochastic evolution equation around the steady hydrodynamic fields and solving the resulting linear problem to obtain the form of the fluctuations \cite{boon91a,zarate06a}. Although this procedure allows to compute the lowest-order correlators of the hydrodynamic fields, information about large fluctuations and higher-order correlators is lost as a consequence of the linearization. Taking into account nonlinear corrections (within the framework of non-linear fluctuating hydrodynamics) can help in understanding the long-time tail behavior of lowest-order correlation functions (the reader can find interesting examples in Refs. \cite{vansaarloos82a,shankar86a,mendl13a,spohn14a,mendl15a,zubarev83a}). However, it has been long recognized that in order to explore arbitrary fluctuations an alternative scheme is needed, one based on the computation of the full stationary probability distribution for the observable of interest. This can be achieved using Macroscopic Fluctuation Theory (MFT) \cite{bertini15a}, which offer a variational formula for the this probability distribution starting from the path integral representation of the fluctuating hydrodynamics of the systems at hand. We refer the interested reader to existing reviews for a general overview of this framework \cite{bertini15a,touchette09a,touchette17a,ellis95a,ellis99a,ellis07a}.

We hence consider a quiescent fluid with an arbitrary initial temperature profile $T(\vrr,0)=\bar T(\vrr)$ at time $t=0$ distributed according to the stationary distribution $P_{st}(\bar T)$, and we are interested in the path probability associated to a particular system trajectory in the mesoscopic phase space spanned by the temperature and current fields, i.e. a trajectory $\{T(\vrr,t'),\vj(\vrr,t')\}_{t'=0}^t$ for all $\vrr\in\Lambda$. This path probability can be obtained from Eqs. (\ref{f4}) and (\ref{f6}) by summing over all noise field realizations $\{\bm\xi\}_0^t$ compatible with trajectory $\{T,\vj\}_{0}^t$, resulting in 
\begin{equation}
P\left(\{T(\vrr,t'),\vj(\vrr,t')\}_{t'=0}^t\right)\propto\exp\left(-\Omega\,\mathcal{I}[T,\vj] \right)\, ,\label{f7}
\end{equation}
with
\begin{equation}
\mathcal{I}[T,\vj]=\int_{0}^t dt'\int_{\Lambda}d\vrr \frac{\big[\vj(\vrr,t')+D(T)\vnabla T(\vrr,t')\big]^2}{2\sigma(T)}
\end{equation}
provided that the current and temperature fields are coupled via the continuity equation, namely
\begin{equation}
\partial_t T(\vrr,t)+\vnabla \cdot \vj(\vrr,t)=0\, .\label{f8}
\end{equation} 

As explained above, we are interested in the statistics of the heat current flowing through the fluid during a long time interval $t$. In particular, we define now the empirical space-and-time-averaged current $\vJJ$ as
\begin{equation}
\vJJ\equiv\frac{1}{t L^d}\int_{0}^t dt'\int_{\Lambda}d\vrr\,\vj(\vrr,t')\, ,
\label{av-current}
\end{equation}
and consider the probability distribution for this observable in the long time and large scale separation (i.e. large $\Omega$) limits. This distribution can be written as the path integral over all possible trajectories of the temperature and current fields which, starting from the fluid's steady state distribution and weighted by (\ref{f7}), are compatible with the required averaged current $\vJJ$ [see Eq. (\ref{av-current})] and the continuity constraint (\ref{f8}),
\begin{widetext}
\be
\mathcal{P}(\vJJ;t)\propto\int \mathcal{D}T\int \mathcal{D}\vj ~P_{st}(\bar T)~P\left(\{T,\vj\}_{0}^t\right) \prod_{t'=0}^t\prod_{\vrr\in\Lambda} \delta\big(\partial_{t'} T+\vnabla \cdot \vj \big)\delta\left(\vJJ t L^d-\int_{0}^t dt'\int_{\Lambda}d\vrr\,\vj \right)\, ,\label{g1}
\ee
where $\delta[\cdot]$ is the Dirac delta function accounting for the different constraints. We can now just use the Laplace representation of the delta function to substitute constraints by Lagrange multipliers, namely a scalar field $\psi(\vrr,t)$ conjugated to the continuity equation constraint and a vector $\vlamb$ conjugated to the current, leading to
\begin{equation}
\mathcal{P}(\vJJ;t)\propto\int \mathcal{D}T\int \mathcal{D}\vj\int \mathcal{D}\psi\int d\vlamb ~P_{st}(\bar T) \exp\big[-\Omega t \cL(T,\vj,\psi,\vlamb;t) \big]\, ,\label{g2}
\end{equation}
with a Lagrangian functional given by
\begin{equation}
\cL(T,\vj,\psi,\vlamb;t)=\frac{1}{t}\int_{0}^t dt'\int_{\Lambda}d\vrr\biggl[ \frac{\left(\vj(\vrr,t')+D(T)\vnabla T\right)^2}{2\sigma(T)} +\psi(\vrr,t') \big[ \partial_{t'} T +\vnabla \cdot \vj \big]+\vlamb\cdot\big[\vJJ-\vj(\vrr,t')\big] \biggr]\, ,\label{g3}
\end{equation}
\end{widetext}
with boundary conditions for the field $\psi$ such that $\psi(\vrr,t)=0$ $\forall \vrr\in\partial\Lambda$, where $\partial\Lambda$ denotes the system boundary. For long times and large scale separation $\Omega$, the probability density function (PDF) of the empirical current obeys a large deviation principle, scaling as $\mathcal{P}(\vJJ;t)\asymp\exp\left[-\Omega t G(\vJJ)\right]$, where the symbol ``$\asymp$'' stands for asymptotic logarithmic equality. This scaling means that this PDF concentrates exponentially fast around its average value, i.e. such that the probability of large fluctuations far from the average decay exponentially with time and $\Omega$. The rate function $G(\vJJ)$ is the \emph{current large deviation function} (LDF), and follows from the previous Lagrangian via a saddle-point calculation in the long-time limit:
\begin{equation}
G(\vJJ)=\lim_{t\rightarrow\infty}\left\lbrace\min_{T,\vj,\psi,\vlamb} \cL(T,\vj,\psi,\vlamb;t)\right\rbrace\,.\label{g5}
\end{equation}
Interestingly, note that $P_{st}(\bar T)$ does not contribute to the LDF since it appears as a sub-dominant term in the long-time limit.

The problem of heat flux statistics can be formulated not in terms of the PDF $\mathcal{P}(\vJJ;t)$ but instead in terms of its cumulant generating function. In particular, we define the \emph{scaled cumulant generating function} (sCGF) as
\begin{equation}\label{scgf}
\mu(\vlamb)\equiv\lim_{t\rightarrow\infty}\frac{1}{t \Omega}\ln{\left\langle e^{t \Omega\vlamb\cdot\vJJ}\right\rangle}\,,
\end{equation}
where the average is defined with respect to the PDF $\mathcal{P}(\vJJ;t)$. The sCGF works as a \emph{dynamical free energy}, and fully characterizes the PDF of the total current $\vJJ$ \cite{tizon-escamilla17a,perez-espigares16a}. The vector $\vlamb$ is conjugated to the averaged current $\vJJ$, in a similar way to the relation between temperature and energy in equilibrium. Indeed, the current $\vJJ_\vlamb$ associated to a given value of the parameter $\vlamb$ is fixed by the relation $\vJJ_\vlamb= \vnabla_\vlamb \mu(\vlamb)$. Moreover, according to G\"artner-Ellis theorem \cite{gartner77a,ellis84a,touchette09a}, the sCGF is directly related to the current LDF via a Legendre-Fenchel transform
\begin{equation}
\mu(\vlamb)=\max_{\vJJ}\left[\vlamb\cdot\vJJ-G(\vJJ)\right]\,.\label{m5}
\end{equation}
The cumulants of the current PDF can be now obtained from the derivatives of the dynamical free energy $\mu(\vlamb)$ evaluated at $\vlamb=0$. In particular, introducing
\begin{equation}
\mu^{(n)}_{(n_1\ldots n_d)} \equiv\left[\frac{\partial^n\mu(\vlamb)}{\partial\lambda_1^{n_1}\ldots\partial\lambda_d^{n_d}}\right]_{\vlamb=0}\, ,
\label{cumul}
\end{equation}
with $\sum_{i=1}^dn_i=n$ and $\lambda_i$ the $i$-th component of the vector $\vlamb$, one can show for $n\le 3$ that $\mu^{(n)}_{(n_1,\ldots,n_d)} = (t \Omega)^{n-1}\left\langle\Delta J_1^{n_1}...\Delta J_d^{n_d}\right\rangle$, with $\Delta J_i\equiv J_i-(1-\delta_{n,1})\langle J_i\rangle$ and $\langle J_i\rangle$ the average current along the $i$-direction. Note that since $\vJJ$ is a space-and-time-averaged current, the cumulants $\mu^{(n)}_{(n_1,\ldots,n_d)}$ are nothing but spatiotemporal integrals of $n$-point correlators of the current field \cite{hurtado14a}.

\section{The most probable path}

We next focus on solving the variational problem defined by (\ref{g5}). This analysis will lead to explicit predictions for the current statistics, as well as to a detailed knowledge of the properties of the optimal (or most probable) path associated to an arbitrary fluctuation. This optimal path follows from the solution $(\TJ,\vjJ,\psiJ,\vlJ)$ of the variational problem (\ref{g5}), and defines the trajectory that the fluid follows in mesoscopic phase space to sustain a long-time current fluctuation. These optimal fields are the solution of the following Euler-Lagrange equations
\begin{eqnarray}
&\displaystyle \partial_t\psiJ=-\frac{\sigma'}{2\sigma^2}\Big(\vjJ^{\,2}-D^2(\vnabla\TJ)^2\Big)-\frac{D}{\sigma}\vnabla\cdot (\vjJ+D\vnabla\TJ)\nonumber\\
&\displaystyle \vjJ+D\vnabla \TJ=\sigma\left(\vnabla\psiJ+\vlJ\right)\label{genPDE}\\
&\displaystyle \partial_t \TJ(\vrr,t)+\vnabla\cdot\vjJ(\vrr,t)=0\nonumber\\
&\displaystyle \vJJ=\frac{1}{t L^d}\int_0^t dt'\int_\Lambda d\vrr\, \vjJ(\vrr,t')\, ,\nonumber
\end{eqnarray}
with $D=D(\TJ)$ and $\sigma=\sigma(\TJ)$, and $\sigma'$ the derivative of $\sigma$ with respect to its argument. As a result, the current LDF takes the form
\begin{equation}
G(\vJJ)=\frac{1}{t}\int_{0}^t dt'\int_{\Lambda}d\vrr \frac{\Big[\vjJ(\vrr,t')+D(\TJ)\vnabla\TJ(\vrr,t')\Big]^2}{2\sigma(\TJ)}\label{g7} \, ,
\end{equation}
in terms of the optimal temperature and current fields. 

The general solution of the spatiotemporal problem (\ref{genPDE}) remains a major challenge in most cases \cite{bertini05a,bertini06a,bertini07a,bertini09a}. However, a powerful conjecture known as additivity principle has been put forward for systems in $d=1$ \cite{bodineau04a,bodineau06a,derrida07a,pilgram03a,jordan04a,shpielberg16a} and recently extended for $d>1$ \cite{perez-espigares16a,tizon-escamilla17a} which strongly simplifies the variational problem at hand. In brief, this additivity principle assumes that, except for initial and final transients of negligible statistical weight, the optimal path associated to a current fluctuation is \emph{time independent}. The validity of this conjecture in open systems has been proved in simulations both for $1D$ stochastic lattice gases \cite{hurtado09c,hurtado10a,gorissen12a,hurtado14a} and $d>1$ driven diffusive models \cite{hurtado11b,perez-espigares16a,tizon-escamilla17a}. We hence adopt the additivity principle here and assume the solutions of Eq. (\ref{genPDE}) to be time independent, i.e. $\TJ(\vrr)$, $\vjJ(\vrr)$ and $\psiJ(\vrr)$. Recalling the boundary conditions for the temperature field described in the previous section, we have that $\TJ(0,\vxperp)=T_0$ and $\TJ(L,\vxperp)=T_1$, together with
\be
\TJ(x,\vxperp +L\,{\bm{\hat a_i}})=\TJ(x,\vxperp)\quad \forall i=2,..,d, \nonumber
\ee
$\forall \vxperp\in[0,L]^{d-1}$, where we have decomposed the position vector $\vrr=(x,\vxperp)$ along the gradient direction ($x$) and all other $(d-1)$ orthogonal directions ($\vxperp$), with ${\bm{\hat a_i}}$ the canonical unit vectors. These boundary conditions correspond to a fluid in contact with two plates at temperatures $T_0$ and $T_1$ at the $x$-boundaries at $x=0$ and $L$, respectively, and periodic boundary conditions on the perpendicular $(d-1)$-subspace. The symmetry of the boundary conditions leads to the natural assumption that the optimal temperature and current fields will exhibit structure only along the $x$-direction, i.e. $\TJ(\vrr)=\TJ(x)$ and $\vjJ(\vrr)=\vjJ(x)$. Together with the additivity principle, this can be shown to imply (see Appendix \ref{app-current} and Refs. \cite{perez-espigares16a,tizon-escamilla17a}) that the optimal current field exhibits a nontrivial structure of the form
\begin{equation}
\vjJ(x,\vxperp)=(J_x,\frac{\sigma(\TJ)}{A(\TJ)}\vJperp)\,, \label{g11}
\end{equation}
with the decomposition $\vJJ=(J_x,\vJperp)$ and
\begin{equation}
 A(\TJ)=\frac{1}{L}\int_0^L dx\,\sigma[\TJ(x)]\, . \label{clausure}
\end{equation}
As a result, the probability $\mathcal{P}(\vJJ;t)$ is completely characterized in terms of the optimal temperature profile $\TJ(x)$. Considering (\ref{genPDE}) and the previous assumptions, the most probable temperature field satisfies the ordinary differential equation \cite{hurtado14a}
\begin{equation}
\left[D(\TJ) \frac{d\TJ}{dx}\right]^2=J_x^2+K\sigma(\TJ)-\left(\frac{\sigma(\TJ)}{A(\TJ)}\right)^2{\bm{J_\perp}^2}\, ,\label{g13}
\end{equation}
where $K$ is an integration constant fixed by the boundary conditions, which are given by $T_0$ and $T_1$. 

In order to proceed, we now need to specify the functional form of the thermal diffusivity and mobility transport coefficients, which completely define the model fluid we will study here. For an incompressible fluid under moderate boundary temperature gradients, the thermal conductivity can be considered a constant of the material, and hence the thermal diffusivity defined above will be a constant, that we take here to be $D=1/2$. Furthermore, in this situation it can be proved using the fluctuation-dissipation theorem that the standard deviation of the fluctuating heat current (which is nothing but the mobility) scales as the local temperature squared \cite{landau13a,landau13b,zarate06a}, so we take $\sigma(T)=T^2$. Indeed, these two transport coefficients define a broadly studied transport model, the Kipnis-Marchioro-Presutti model of heat conduction \cite{kipnis82a} which, as we see here, captures the heat transport properties of a quiescent incompressible fluid. With these prescriptions, the differential equation (\ref{g13}) boils down to
\begin{equation}
\frac{d\TJ}{dx}=\pm 2\left[J_x^2+K\TJ^2-\frac{\TJ^4}{A^2}\vJperp^2 \right]^{1/2}\, ,\label{Tder}
\end{equation}
where we have fixed $L=1$ for simplicity. This equation can be solved in terms of Jacobi inverse elliptic functions (see Appendix \ref{app-temperature}), leading to the following reduced optimal temperature field
\begin{equation}
\tau(x)\equiv \frac{\TJ(x)}{T_1}=\frac{\cn\Big[-F_0+(F_1+F_0)x;k\Big]}{\cn(F_1;k)}\, , \label{p1}
\end{equation}
where $\cn(u;k)$ is the cosine-amplitude Jacobi function with modulus $k$ \cite{gradshteyn07a,byrd71a}. The value of the constant parameters $F_{0,1}$, as well as the modulus, are fixed by the boundary conditions and the closure equation (\ref{clausure}), namely
\begin{eqnarray}
Q_1&=&\frac{\sqrt{1-k^2}(F_1+F_0)}{2 ~\cn(F_1;k)}\,,\label{p2}\\
Q_\perp&=&\frac{E_1+E_0-(1-k^2)(F_1+F_0)}{2k ~\cn(F_1;k)}\,,\label{p3}\\
\tau_0&=&\frac{\cn(F_0;k)}{\cn(F_1;k)}\,,\label{p4}
\end{eqnarray}
where we have defined $Q_1=|J_x|/T_1$, $Q_\perp=|\vJperp|/T_1$, $\tau_0=T_0/T_1$ and $E_{0,1}=E(\am(F_{0,1};k);k)$ with $\am(u;k)$ the amplitude Jacobi function and $E(\theta;k)$ the Jacobi integral of the second kind \cite{gradshteyn07a,byrd71a}. Note that, assuming without loss of generality that $T_0\ge T_1$ so $\tau_0\geq 1$, we have that  $F_0\in[-{\bf K}(k),{\bf K}(k)]$, $F_1\in [\cn^{-1}(\tau_0^{-1};k),{\bf K}(k)]$ and $k\in[0,1]$, with $F_1\geq F_0$ and ${\bf K}$ the Jacobi complete elliptic integral of the first kind (using the notation of Gradshteyn and Ryzhik \cite{gradshteyn07a}). In this way, once the physical variables $Q_1$, $Q_\perp$ and $\tau_0$ are fixed, we can obtain $F_0$, $F_1$ and $k$ from Eqs. (\ref{p2})-(\ref{p4}). Note also that, for a fixed value of the external gradient parameter $\tau_0$, one can solve Eq. (\ref{p4}) to obtain
\be
F_1=\cn^{-1}\Big[\frac{1}{\tau_0}\cn(F_0;k);k\Big] \, ,
\ee
Therefore, substituting $F_1$ into Eqs. (\ref{p2}) and (\ref{p3}), we conclude that $Q_1$ and $Q_\perp$ are just functions of $F_0$ and $k$.

\section{Scaling, structure and universality of the optimal path}\label{sect4}

As shown above, the most probable reduced temperature profile $\tau(x)$ is a continuous positive function written in terms of $\cn(u;k)$, an even and periodic function of its argument $u=-F_0+(F_1+F_0)x$. Indeed, the cosine-amplitude Jacobi function presents only one positive maximum located at $u=0$ \cite{gradshteyn07a,byrd71a}, i.e., $x_{max}=F_0/(F_0+F_1)$, which implies that the optimal temperature field (defined in the spatial interval $x\in[0,1]$) exhibits at most two possible typical behaviors: (1) a \emph{single-maximum profile} for $F_0>0$, or (2) a \emph{monotonously decreasing profile} for $F_0<0$. The values of $Q_1$ and $Q_\perp$ where the crossover happens can be found by setting $F_0=0$ and $F_1=\cn^{-1}(1/\tau_0;k)$ on Eqs. (\ref{p2}) and (\ref{p3}), and are a function of the modulus $k\in[0,1]$ and the external gradient parameter $\tau_0$. This condition defines a limiting curve in the $Q_1-Q_\perp$ plane for each $\tau_0$ separating both behaviors.

\begin{figure}
\includegraphics[width=8cm,clip]{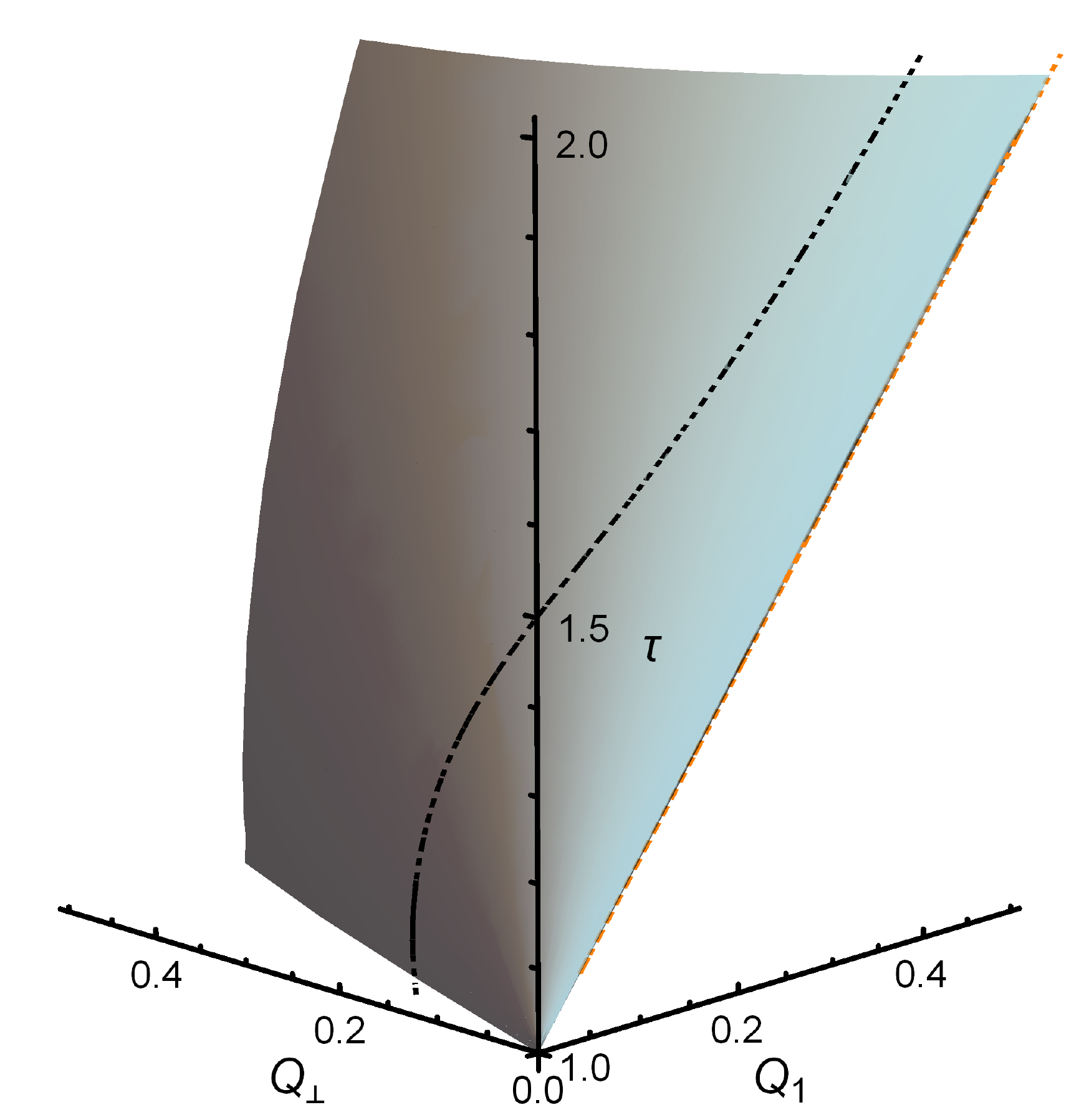}
\caption{Surface defined by the set of points $(\tau_0,Q_1,Q_\perp)$ with fixed modulus $k=0.9$. Points in this surface have the same scaling form of the associated optimal reduced temperature field $\tau(x)$ except for a linear transformation of the $x$-coordinate and a suitable amplitude factor, see Eq. (\ref{scaling}). The black dashed line shows the set of points in this surface with the aditional constraint $F_1+F_0=0.4$. Reduced optimal profiles along this curve present the same functional structure except for only a translation of the $x$-coordinate. The orange dashed line represents the stationary current values given by $(\tau_0,Q_1^{st},Q_\perp^{st})=(\tau_0,(\tau_0-1)/2,0)$.}
\label{fig1}
\end{figure}

\begin{figure*}
\begin{center}
\includegraphics[width=18cm,clip]{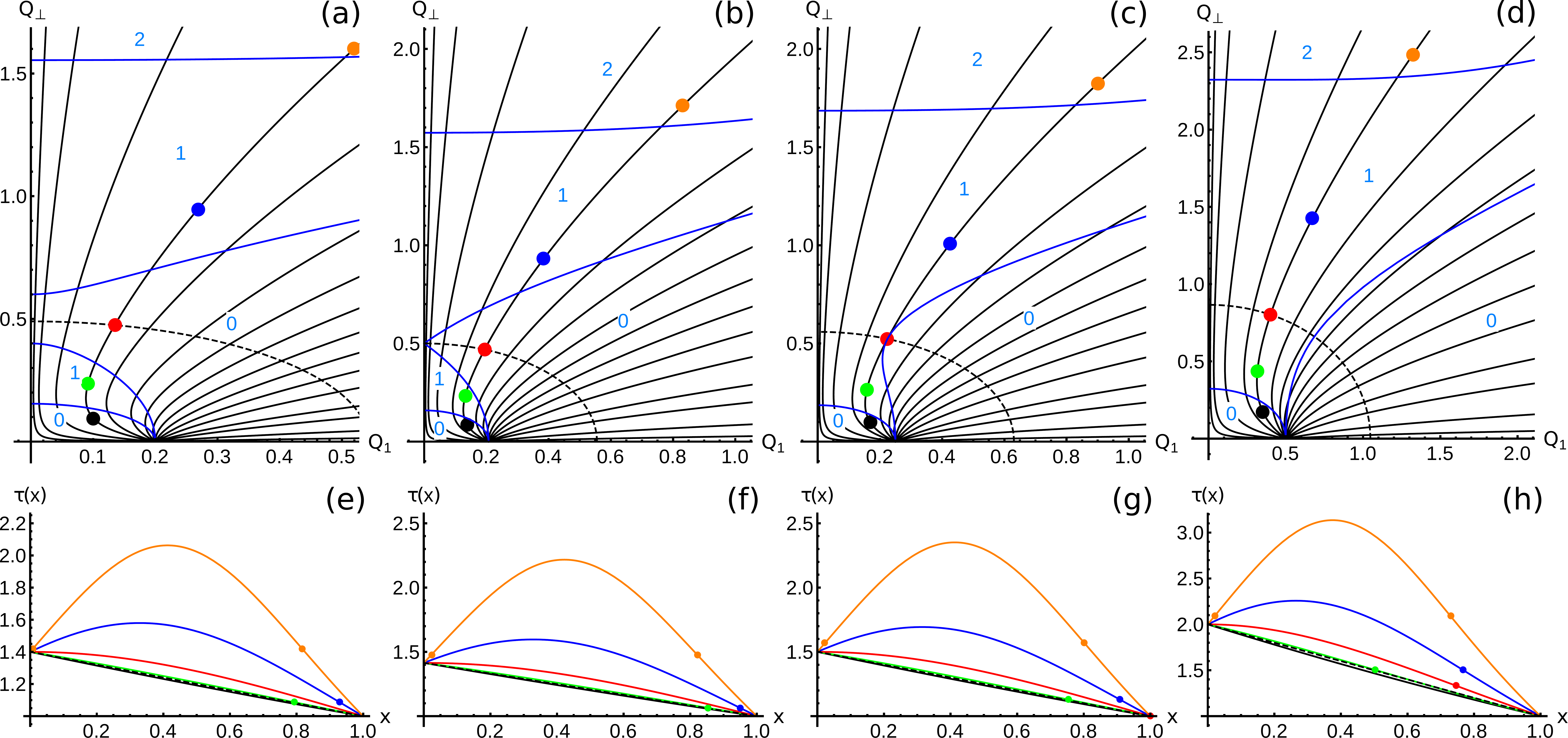}
\end{center}
\caption{Top row (a-d): Current fluctuations exhibiting the same scaling form of the optimal reduced temperature profile. Each black solid line represents a uniparametric family of solutions $(Q_1(F_0),Q_\perp(F_0))$ of Eqs. (\ref{p4}) and (\ref{p3}) with varying $F_0$ and fixed $k$ which share the same scaling form of the optimal profile. Each panel includes curves for $k=0.001$, $0.01$, $0.05$, $0.1$, $0.2$, $0.3$, $0.4$, $0.5$, $0.6$, $0.7$, $0.8$, $0.9$, $0.95$, $0.99$, $0.999$, $0.9999$ (displayed counterclockwise). Each panel corresponds to a fixed external gradient parameter $\tau_0$, with $\tau_0=1.4,\sqrt{2},1.5,$ and $2$ from left to right. The dashed black line in each panel represents the crossover between monotonous (below the dashed line, $F_0<0$) and non-monotonous, single-maximum profiles (above the dashed line, $F_0>0$). Blue lines separate regions of profiles with $0$, $1$ and $2$ inflection points. Bottom row (e-h): Optimal reduced temperature profiles associated to the different highlighted dots in upper panels. The dashed lines represents the stationary profile in each case, while the dots locate the corresponding inflection points (if any).}
\label{fig2}
\end{figure*}

Interestingly, Eqs.~(\ref{p2})-(\ref{p4}) lead to a one-to-one correspondence between the set of physical variables $(\tau_0,Q_1,Q_\perp)$ and the parameters $(k,F_0,F_1)$. The Jacobi-cosinus function $cn(u;k)$ defining the most probable temperature profile (\ref{p1}) is just a linear function of space, $u=-F_0+(F_1+F_0)x$, with constants fixed by $(\tau_0,Q_1,Q_\perp)$, while the modulus $k$ captures the particular functional dependence on $u$ [e.g. $\cn(u;k=0)=\cos u$ while $\cn(u;k=1)=\text{sech}~u$]. In this way, the modulus $k$ \emph{parametrizes} in a natural way the topology of the optimal temperature field: all optimal profiles with the same modulus $k$ share the same functional structure (after a linear transformation of the $x$-coordinate and a suitable amplitude factor). Therefore there exist a surface in $(\tau_0,Q_1,Q_\perp)$-space, defined by the constraint on constant $k$, whose optimal reduced temperature profiles follow the scaling function
\begin{equation}
\tau(x)=A(\tau_0,Q_1,Q_\perp) \cn(u;k) \, , \quad  u=-F_0+(F_0+F_1)x \, .
\label{scaling}
\end{equation}
This defines a universal scaling behavior for the optimal temperature fields responsible for different current fluctuations in the quiescent incompressible fluid. Note in particular that the above scaling implies the existence of optimal profiles associated to different values of the external gradient parameter $\tau_0=T_0/T_1$ with the same functional form. Figure \ref{fig1} shows an example of the surface of points $(\tau_0,Q_1,Q_\perp)$ having the same value of $k=0.9$ and hence the same scaling behavior. We note that these surfaces are analytic at all points. Finally, one can define a stronger universal scaling by demanding that not only the modulus $k$ is fixed, but also the slope $F_0+F_1$ of the linear map in the scaling function (\ref{scaling}). This additional constraint defines a curve within the $(\tau_0,Q_1,Q_\perp)$-surface of constant-$k$ along which the optimal temperature field for a heat current fluctuation has the same functional form except for a \emph{translation} along the $x$-coordinate (see the black dashed line in Fig. \ref{fig1}).

The top row in Fig. \ref{fig2} presents with black solid lines different families of current fluctuations which share the same scaling form of the optimal temperature field (i.e. have the same value of the modulus $k$) for different values of the external gradient parameter $\tau_0$. Note that these curves are parametrized by $F_0$ for each fixed $\tau_0$. Remarkably, we observe that all curves of current fluctuations converge to the stationary value $(Q_1^{st},Q_\perp^{st})=[(\tau_0-1)/2,0]$ when $F_0\rightarrow-{\bf K}(k)$, implying that around the nonequilibrium stationary state \emph{all} family members have monotonous temperature profiles ($F_0<0$) and contribute to the fluctuating behavior of $\vJJ$'s with a probability whose value will be study in the next section. In particular, we emphasize that all possible scaling structures of the optimal temperature profile are present when we consider infinitesimally small fluctuations around the steady-state current, the dominant family being determined by the \emph{orientation} of the infinitesimal current fluctuation vector.

Finally, we have also studied the convexity properties of the optimal temperature field by analyzing in detail the form of its second derivative, finding profiles with $0$, $1$ or $2$ inflection points. This rich phenomenology is also displayed in Fig. \ref{fig2} (top row), where we show for varying $\tau_0$ the regions corresponding to profiles with different numbers of inflection points (blue solid lines and numbers). In addition, the particular shape of the most probable temperature fields for different values of $(\tau_0,Q_1,Q_\perp)$ signaled with points in the upper panels is also shown; see bottom row in Fig. \ref{fig2}. Important features to note here are the transition from \emph{monotonous} to \emph{single-maximum} profiles as the distance to the stationary state is increased (measured in terms of the current), as well as the change in the number of inflection points appearing in each one (identified with a dot). The evolution of the number of inflection points as we move away from the stationary current is non-trivial, and we notice the reentrant behavior of the curve delimiting the regime of current fluctuations whose optimal profiles have no inflection points. This reentrance changes as the external gradient parameter $\tau_0$ is varied, disappearing for large enough $\tau_0$. It is also interesting to stress that the curves delimiting the number of inflection points intersect with the curves defining the different scaling profile families for constant $k$ (see top panels in Fig. \ref{fig2}) meaning that profiles within the same scaling family can exhibit a variable number of inflection points despite having the same overall functional form.

\section{Heat current statistics}\label{sect5}

Once the optimal temperature profiles have been determined, we are in position to study in detail the probability density function $\mathcal{P}(\vJJ;t)$ of the fluid's empirical heat current $\vJJ$. As shown in Sec. \ref{sectmodel}, the PDF $\mathcal{P}(\vJJ;t)$ obeys a large deviation principle for long times of the form $\mathcal{P}(\vJJ;t)\asymp\exp\left[-t\Omega G(\vJJ)\right]$, which defines the current LDF $G(\vJJ)$. The MFT equations lead to a variational problem for $G(\vJJ)$, which can be written in terms of the optimal temperature and current fields as shown in Eq. (\ref{g7}). As a result, using the additivity principle \cite{bodineau04a} and taking into account the structure of the optimal temperature fields (\ref{p1}) and its relation with the optimal heat current (\ref{g11}), we arrive at the following expression for the current LDF
\begin{widetext}
\begin{equation}
G(\vJJ)=\frac{J_x}{2}\left(\frac{1}{T_0}-\frac{1}{T_1}\right)+\frac{1}{8}(F_0+F_1)^2  + \frac{1}{4}(F_0+F_1) \left(\frac{\sn(F_1;k)~\dn(F_1;k)}{\cn(F_1;k)}+\frac{\sn(F_0;k)~\dn(F_0;k)}{\cn(F_0;k)}-E_0-E_1\right)\, ,
\label{gtot}
\end{equation}
\end{widetext}
written in terms of the parameters $(k,F_0,F_1)$ linked to the physical variables $(\tau_0,Q_1,Q_\perp)$ via Eqs. (\ref{p2})-(\ref{p4}). From this expression, it is easy to check that the Gallavotti-Cohen fluctuation theorem \cite{evans93a,evans94a,gallavotti95a,gallavotti95b,kurchan98a,lebowitz99a}, relating the probability of an arbitrary current fluctuation $\vJJ$ with its time-reversed current $-\vJJ$, holds in this case, namely
\begin{equation}
G(\vJJ)-G(-\vJJ)=2{\bm \epsilon}\cdot \vJJ = 2|{\bm \epsilon}| J_x\, ,\label{Galla}
\end{equation} 
where ${\bm \epsilon}=\frac{1}{2}\left(T_0^{-1}-T_1^{-1}\right)\hat{x}$ is the nonequilibrium driving force (with $\hat{x}$ the unit vector along the gradient direction), related to the rate of entropy production in the nonequilibrium fluid appearing as a consequence of the boundary temperature gradient. Moreover, the symmetry of the problem implies that the LDF also satisfies $G(J_x,\vJperp)=G(J_x,-\vJperp)$ $\forall J_x,\vJJ_\perp$.

Interestingly, for $1D$ driven diffusive systems with a quadratic mobility $\sigma(T)$, it has been recently shown \cite{bodineau06a,derrida09a} that the current LDF has a simple expression in terms of a non-trivial scaling parameter. A natural question is then whether the current LDF derived here for a $d$-dimensional quiescent fluid with a quadratic mobility $\sigma(T)=T^2$ obeys a similar scaling picture. As far as we know, there exist no such a simple scaling formulation for fluctuations of the \emph{vectorial} current in terms of a single, well-defined parameter, though still there can exist an analogous transformation when considering fluctuations of the empirical current restricted to the direction of the boundary gradient. We believe that the main reason of such lack of simple scaling resides in the non-trivial structure adopted by the most probable current vector field [see Eq. (\ref{g11})] a direct consequence of the Weak Additivity Principle \cite{perez-espigares16a,tizon-escamilla17a}.

\begin{figure}
\begin{center}
\includegraphics[width=8cm,clip]{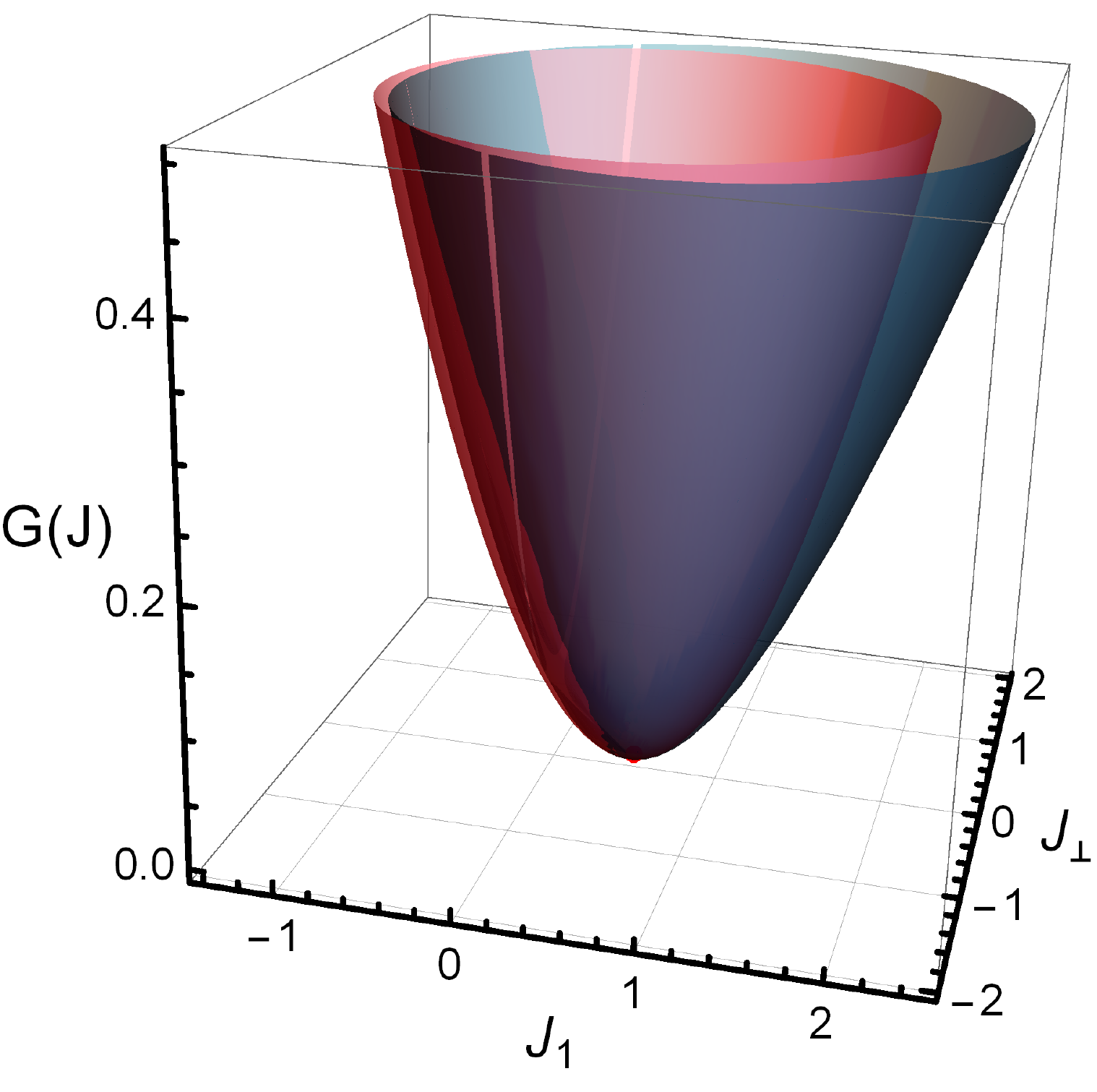}
\end{center}
\caption{The dark (outer) surface represents the exact current LDF $G(\vJJ)$ for $T_0=2$ and $T_1=1$ (so $\tau_0=2$), see Eq. (\ref{gtot}). The red (inner) surface corresponds to Gaussian approximation $G_{\text{gauss}}(\vJJ)$ around the stationary state for the same parameters, see Eq. (\ref{Ggauss}). The red point at the bottom represents the stationary state.}
\label{fig4}
\end{figure}

To better understand the fluid's heat current statistics, it is interesting to analyze the behavior of $G(\vJJ)$ in two opposing limits, i.e. for small current fluctuations around the stationary state defined by ${\bf{J_{st}}}=\left(J_x^{st}=T_1(\tau_0-1)/2,{\bm{J_\perp^{st}}}=0\right)$, and its behavior in the far tails of the distribution. In the first case, by expanding $G(\vJJ)$ around ${\bf{J_{st}}}$ keeping only up to second order contributions, the current LDF can be approximated by (see Appendix \ref{app3})
\begin{widetext}
\begin{eqnarray}
G({\bm{\tilde{Q}}})&=&G_{\text{gauss}}({\bm{\tilde{Q}}})-\frac{3(\tilde Q_1^2+\tilde Q_\perp^2)}{2(1+\tau_0+\tau_0^2)}\biggl[\frac{2(\tau_0-1)(4+7\tau_0+4\tau_0^2)}{5(1+\tau_0+\tau_0^2)^2}\tilde Q_1\nonumber\\
&+&\frac{9}{175(1+\tau_0+\tau_0^2)^4}\biggl(-5(4+2\tau_0-30\tau_0^2-57\tau_0^3-30\tau_0^4+2\tau_0^5+4\tau_0^6)(\tilde Q_1^2+\tilde Q_\perp^2)\nonumber\\
&+&2(\tau_0-1)^2(16+61\tau_0+91\tau_0^2+61\tau_0^3+16\tau_0^4)(\tilde Q_\perp^2-\tilde Q_1^2)\biggr)
+O({\bm{\tilde{Q}}}^3)\biggr]\, ,\label{Gstat}
\end{eqnarray}
where we have introduced an excess reduced current vector ${\bm{\tilde{Q}}}=(\tilde{Q}_1,\tilde{Q}_\perp)$, with the definitions $\tilde{Q}_1\equiv Q_1-\vert J_x^{st}\vert/T_1$, $\tilde{Q}_\perp\equiv Q_\perp-\vert\bm{J_\perp^{st}}\vert/T_1$, and where
\begin{equation}
G_{\text{gauss}}({\bm{\tilde Q}})=\frac{3(\tilde Q_1^2+\tilde Q_\perp^2)}{2(1+\tau_0+\tau_0^2)}\,  \label{Ggauss}
\end{equation}
captures the Gaussian fluctuations around the steady state expected from the central limit theorem. In Fig. \ref{fig4} we represent the exact $G(\vJJ)$ of Eq. (\ref{gtot}) (dark outer surface) for $\tau_0=2$, together with the Gaussian part of the expansion (\ref{Gstat}), $G_{\text{gauss}}$, (red inner surface). We stress here the non-Gaussian, asymmetric structure of the exact $G(\vJJ)$, which can be however approximated by a deformed Gaussian on both axis at least for moderate current fluctuations. The dominant corrections of the optimal reduced temperature field beyond the steady-state (linear) profile can be also computed to first order in ${\bm{\tilde Q}}$, leading to
\begin{equation}
\tau(x)=\tau_0-x(\tau_0-1)+\frac{2(\tau_0-1)}{1+\tau_0+\tau_0^2}x(1-x)(1+2\tau_0-x(\tau_0-1))\tilde Q_1+O({\bm{\tilde{Q}}}^2)\, ,
\end{equation}
i.e. a polynomial deformation of the linear stationary profile.

We are also interested on the leading behavior of $G(\vJJ)$ for currents far from stationary state behavior. This can be studied in detail by focusing on two different limits, namely $\left(\vert J_x\vert \gg J_x^{st},\vJperp=0\right)$ and $\left(J_x=0,\vert \vJperp\vert\gg 0\right)$. The corresponding expansion is performed in Appendix \ref{app3}, and leads to
\begin{eqnarray}
G(J_x,0)&=&\frac{J_x}{T_0}-\frac{\pi^2}{8}+\frac{\pi^2(1+\tau_0)T_1}{16J_x}+O(J_x^{-2})\, \qquad \text{for } \vert J_x\vert \gg J_x^{st} \\
G(0,\vert \vJperp\vert)&=&\frac{1}{8}\ln\left(\frac{4\vert {\vJperp}\vert^2T_1^3}{T_0}\right)\left[\ln\left(\frac{4\vert {\vJperp}\vert^2T_1^3}{T_0}\right)+\frac{1+\tau_0^2}{T_1^2 \vert {\vJperp}\vert^2}+O(\vert {\vJperp}\vert^{-4})\right]\, \qquad \text{for } \vert \vJperp\vert\gg 0.
\end{eqnarray}
This implies in particular that large current fluctuations along the gradient direction decay exponentially in the current, rather than in a Gaussian manner as a naive central-limit analysis would suggest. More interestingly, the statistics of large current fluctuations \emph{orthogonal} to the thermal gradient exhibit logarithmic behavior, which makes these rare fluctuations much more probable than anticipated within the Gaussian approximation. This interesting behavior points out once again to the complex analytic behavior of the heat current LDF, in contrast with the apparent smooth and simple structure shown in Fig. \ref{fig4} for moderate current fluctuations.

With the aim of computing the first few cumulants of the current distribution, we calculate now the scaled cumulant generating function (sCGF) $\mu(\lambda)$ of the current distribution, see Eq. (\ref{scgf}) and Sec. \ref{sectmodel}. Indeed, considering the form of $G(\vJJ)$ near the stationary state [Eq. (\ref{Gstat})] and the Legendre duality between $\mu(\lambda)$ and $G(\vJJ)$ [Eq. (\ref{m5})] the sCGF can be expanded as
\begin{eqnarray}
\mu(\vlamb)&=&\frac{1}{2}(T_0-T_1)\lambda_1+(\lambda_1^2+{\bm{\lambda_\perp}^2})\biggl[\frac{1}{6}(T_0^2+T_0T_1+T_1^2)+\frac{1}{45}(T_0-T_1)(4T_0^2+7T_0T_1+4T_1^2)\lambda_1\nonumber\\
&+&\frac{9}{1890}(12T_0^4+8T_0^3T_1-5T_0^2T_1^2+8T_0T_1^3+12 T_1^4)\lambda_1^2\nonumber\\
&+&\frac{1}{1890}(44T_0^4+76 T_0^3T_1+75 T_0^2T_1^2+76 T_0T_1^3+44T_1^4){\bm{\lambda_\perp}^2}
\biggr]+O(\vlamb^5)\, ,\label{m7}
\end{eqnarray}
where we have decomposed $\vlamb=(\lambda_1,\vlamb_\perp)$ along the gradient ($\lambda_1$) and all orthogonal ($\vlamb_\perp$) directions. We are now in position to compute the lower-order cumulants by differentiating with respect to the components of the $\vlamb$-vector, see Eq. (\ref{cumul}). The first derivatives yield the steady state value of the current components, $\langle J_x\rangle =J_x^{st}=(T_0-T_1)/2$ and  $\langle J_\alpha\rangle =0$, $\forall \alpha\neq x$. The next few cumulants for arbitrary boundary temperatures $T_0$ and $T_1$ compatible with the perturbation expansions ($\tau_0>1$) can be written as
\begin{eqnarray}
\lim_{t\rightarrow\infty}t\Omega\langle(J_s-J_s^{st})^2\rangle&=&\frac{1}{3}(T_0^2+T_0T_1+T_1^2)\nonumber\\
\lim_{t\rightarrow\infty}(t\Omega)^2\langle(J_x-J_x^{st})^3\rangle&=&\frac{2}{15}(T_0-T_1)(4T_0^2+7T_0T_1+4T_1^2)\nonumber\\
\lim_{t\rightarrow\infty}(t\Omega)^2\langle(J_x-J_x^{st})J_\alpha^2\rangle&=&\frac{2}{45}(T_0-T_1)(4T_0^2+7T_0T_1+4T_1^2)\nonumber\\
\lim_{t\rightarrow\infty}(t\Omega)^3\langle(J_x-J_x^{st})^4\rangle&=&\frac{4}{35}(12T_0^4+8T_0^3T_1-5T_0^2T_1^2+8T_0T_1^3+12T_1^4)\nonumber\\
\lim_{t\rightarrow\infty}(t\Omega)^3\langle(J_x-J_x^{st})^2J_\alpha^2\rangle&=&\frac{2}{945}(76T_0^4+74 T_0^3T_1+15T_0^2T_1^2+74T_0T_1^3+76T_1^4)\nonumber\\
\lim_{t\rightarrow\infty}(t\Omega)^3\langle J_\alpha^4\rangle&=&\frac{4}{315}(44T_0^4+76 T_0^3T_1+75T_0^2T_1^2+76T_0T_1^3+44T_1^4)\nonumber\\
\lim_{t\rightarrow\infty}(t\Omega)^3\langle J_\alpha^2J_\beta^2\rangle&=&\frac{4}{945}(44T_0^4+76 T_0^3T_1+75T_0^2T_1^2+76T_0T_1^3+44T_1^4)\, ,\label{m10}
\end{eqnarray}
\end{widetext}
where $s\in[1,d]$ and $\alpha\neq\beta$, with $\alpha,\beta\in[2,d]$ corresponding to any pair of different coordinates in the subspace orthogonal to $x$. Interestingly, remarkable relations between different cumulants can be now derived from (\ref{m10}). In particular
\begin{eqnarray}
3\lim_{t\rightarrow\infty}t^2\langle(J_x-J_x^{st})J_\alpha^2\rangle&=&\lim_{t\rightarrow\infty}t^2\langle(J_x-J_x^{st})^3\rangle\nonumber\\
3\lim_{t\rightarrow\infty}t^3\langle J_\alpha^2J_\beta^2\rangle&=&\lim_{t\rightarrow\infty}t^3\langle J_\alpha^4\rangle\, .\label{m11}
\end{eqnarray}
Whether these relations are a particular result restricted to this model fluid, or rather they reflect a deeper underlying symmetry, remains unknown at this point. A possible origin for these interesting cumulant relations is the Gallavotti-Cohen symmetry \eqref{Galla} stemming from the reversibility of microscopic dynamics. In Ref. \cite{andrieux07a}, Andrieux and Gaspard showed that the Gallavotti-Cohen fluctuation theorem leads to interesting relations among the linear and non-linear response coefficients of the current at arbitrary orders. However, the relations \eqref{m11} are not restricted to the linear-response regime, being valid for arbitrary driving; moreover, these relations link cumulants of the same order (unlike the Andrieux-Gaspard hierarchies). We thus conclude that, as far as we know, the Gallavotti-Cohen symmetry is not uniquely responsible of the observed relations, but might play a key role. In particular, we believe that the set of equations \eqref{m11} reflects also the particular structure of the optimal fields for this problem.

\section{Conclusion}

In summary, we have delved into the heat current statistics of an incompressible quiescent $d$-dimensional fluid subject to a boundary temperature gradient in one direction. This analysis has been carried out within the framework of fluctuating hydrodynamics, using tools borrowed from large deviation theory and macroscopic fluctuation theory. This framework provides powerful techniques to determine the full heat current probability distribution, based on the computation of the most probable trajectories and the current LDF. In this way, under the well-established additivity conjecture (which considers the optimal paths sustaining atypical values of the current to be time-independent), we have determined the explicit form of the most probable temperature fields. We have analyzed their topological properties as a function of the external baths temperatures $(T_0,T_1)$ and the desired empirical current $\vJJ$, defining different regimes where temperature profiles exhibit varying behaviors. Interestingly, our solution to the fluctuation problem shows that optimal temperature fields can be naturally classified in an infinite set of curves, each set sharing the same mathematical structure, parametrized in terms of the modulus $k$ of a Jacobi inverse elliptic function. 

Such characterization of the optimal temperature fields opens the door to the computation of the full heat current probability distribution, as codified in the current LDF. In particular, we have obtained the exact analytical form of the heat current LDF, analyzing its behavior both for small fluctuations around the nonequilibrium steady state, and in the far tails of the distribution. We observe that near the stationary state corrections to Gaussian behavior are small, and the heat current distribution can be well approximated by a deformed Gaussian along all directions. On the other hand, the behavior of current LDF for large values of the current is far more complex, pointing out to the intricateness of fluctuations far from equilibrium. In particular, we find logarithmic tails in the current LDF for large fluctuations orthogonal to the thermal gradient, showing that these fluctuations are far more probable than previously anticipated. Finally, reformulating the statistical problem in terms of the associated cumulant generating function, we have obtained analytic formulas for the first few cumulants of the heat current distribution. These results allow us to find new relations between some of these cumulants, which imply integral relations between different correlators of the heat current field. This finding opens the door to further experimental research to test these results, as the lower-order cumulants of both the empirical heat current and the temperature can be readily measured in actual experiments \cite{katsaros77a,gollub91a,hayashi89a,coppin86a,hall00a,Krishnamurthy16a,garnier09a,zhao09a,mcphee92a,vickers97a,champagne77a}. Prospective experimental systems where to look for these effects are for instance nonequilibrium fluids confined at the nanoscale, where fluctuations play a dominan role, as well as turbulent media or even fusion plasmas characterized by strong temperature gradients along magnetic field lines.

\begin{acknowledgments}
We acknowledge financial support from Ministerio de Economía y Competitividad project FIS2017-84256-P and Ministerio de Educación, Cultura y Deporte fellowship with reference FPU13/05633.
\end{acknowledgments}

 \bibliography{/Users/phurtado/Dropbox/PAPERS/BIBLIOGRAPHY/referencias-BibDesk-OK}

\onecolumngrid
\appendix

\section{Optimal current field}\label{app-current}
In this appendix we prove a relevant property satisfied by the most probable path sustaining an atypical value of the space-and-time-averaged heat current.
\vspace{0.25cm}

\noindent \emph{\it {\bf Property:} Consider the following boundary conditions for the optimal temperature field
\begin{equation}
\TJ(0,\vxperp)=T_0\quad ;\quad \TJ(L,\vxperp)=T_1 \quad ;\quad \TJ(x,\vxperp +L{\bm{\hat a_i}})=\TJ(x,\vxperp)\quad \forall i=2,..,d \,, 
\end{equation}
where we write $\vrr=(x,\vxperp)$ with $x\in[0,L]$ and  $\vxperp\in[0,L]^{d-1}$, where ${\bm{\hat a_i}}$ are the canonical unit vectors. If $\TJ(\vrr)=\TJ(x)$ and $j_{\vJJ,x}(\vrr)=j_{\vJJ,x}(x)$, with $j_{\vJJ,x}$ the component of the current in the $x$-direction, then the most probable current is of the form
\begin{equation}
\vjJ(x,\vxperp)=(J_x,\frac{\sigma(\TJ)}{A(\TJ)}\vJperp)\quad ; \quad A(\TJ)=\frac{1}{L}\int_0^L dx\, \sigma[\TJ(x)]\, .\label{app2-1}
\end{equation}
}

\noindent {\bf Proof:} The additivity principle conjectures that the optimal path associated to a current fluctuation is time-independent. Under this hypothesis, the set of coupled equations (\ref{genPDE}) derived in the main text transforms into
\begin{eqnarray}
&\displaystyle \frac{\sigma'}{2\sigma^2}\left(\vjJ\,^2-D^2(\vnabla\TJ)^2\right)+\frac{D}{\sigma}\vnabla(D\vnabla\TJ)=0\label{genPDE1}\\
&\displaystyle \vjJ+D\vnabla \TJ=\sigma_\vJJ\left(\vnabla\psiJ+\vlamb_\vJJ\right)\label{genPDE2}\\
&\displaystyle \vnabla\cdot\vjJ=0\label{genPDE3}\\
&\displaystyle \vJJ=\frac{1}{\Lambda}\int_\Lambda d\vrr\,\vjJ\, ,\label{genPDE4}
\end{eqnarray}
where $\TJ=\TJ(\vrr)$, $\vjJ=\vjJ(\vrr)$, $D=D(\TJ)$ and $\sigma=\sigma(\TJ)$. Writing Eq. (\ref{genPDE2}) in components and assuming the field $\psiJ$ to be twice continuously differentiable in its spatial domain, the following general property can be proved in general \cite{tizon-escamilla17a,perez-espigares16a}:
\begin{equation}
\partial_\alpha\left(\frac{j_{\vJJ,\beta}}{\sigma}\right)=\partial_\beta\left(\frac{j_{\vJJ,\alpha}}{\sigma}\right);\quad\alpha,\beta\in[1,d],\label{pr1}
\end{equation}
where $j_{\vJJ,\gamma}$ is the $\gamma$ component of the vector field $\vjJ(\vrr)$, and $\partial_\gamma$ the spatial derivative with respect to the coordinate $x_\gamma$. Furthermore, one can easily realize that Eq. (\ref{genPDE1}), together with $\TJ(\vrr)=\TJ(x)$, leads to
\begin{equation}
\vjJ\,^2=j_{\vJJ,x}^2+\sum_{\alpha\neq x} j_{\vJJ,\alpha}^2=F(x)\label{jsquare},
\end{equation}  
with $F(x)$ a function depending only on coordinate $x$. Therefore, assuming $j_{\vJJ,x}(\vrr)=j_{\vJJ,x}(x)$, we obtain from Eq. (\ref{pr1}) that $j_{\vJJ,\alpha}=C_\alpha(\vxperp)\sigma$, where the $\alpha$ subscript refers to all coordinates orthogonal to $x$, and $C_\alpha(\vxperp)$ is a function depending (at most) on the orthogonal coordinates $\vxperp$. Using this expression in Eqs. (\ref{pr1}), (\ref{jsquare}), and (\ref{genPDE3}) we find
\begin{eqnarray}
&\displaystyle\partial_\alpha C_\beta=\partial_\beta C_\alpha\label{Ccond1}\\
&\displaystyle\sum_\alpha C_\alpha^2=R\label{Ccond2}\\
&\displaystyle\sum_\alpha\partial_\alpha C_\alpha=W\label{Ccond3},
\end{eqnarray}
respectively, with $R$ and $W$ two constants. At this point, differentiating Eq. (\ref{Ccond2}) with respect to $x_\beta$, and taking into account Eq. (\ref{Ccond1}), it can be shown that
\begin{equation}
\sum_\alpha C_\alpha\partial_\alpha C_\beta=0.
\end{equation}
Differentiating again with respect to $x_\beta$ and summing over all $\beta$-coordinates, together with Eqs. (\ref{Ccond1}) and (\ref{Ccond3}), we arrive at
\begin{equation}
\sum_\alpha\sum_\beta\left(\partial_\alpha C_\beta\right)^2=0,
\end{equation}
which implies that $C_\beta(\vxperp)$ should be a constant. As a result, the most probable current field is of the form $\vjJ(\vrr)=\left(j_{\vJJ,x}(x),{\bm{C_\perp}}\sigma\right)$ which, considering Eqs. (\ref{genPDE3}) and (\ref{genPDE4}), finally leads to (\ref{app2-1}), as we want to prove. Note that in dimension $d=2$ it can be proved that the optimal current field $\vjJ$ is of the form (\ref{app2-1}) by only hypothesizing $\TJ=\TJ(x)$.

\section{Optimal temperature field}\label{app-temperature}

In this Appendix we determine the analytical expression for the most probable temperature field. Moreover, we will exhaustively analyze its mathematical properties in order to better characterize the statistics of the heat current in our incompressible quiescent model fluid. For our particular model fluid, the optimal temperature field associated to a space-and-time-averaged heat current fluctuation $\vJJ$ is the solution of the following differential equation (see main text):
\begin{equation}
\frac{d\TJ}{dx}=2s\left[J_x^2+K\TJ^2-\frac{\TJ^4}{A^2}{\bm{J_\perp}^2} \right]^{1/2}\, ,\label{Tderapp}
\end{equation}
with $s=\pm 1$ and where, for simplicity, we have fixed $L=1$ without loss of generality. This expression can be rewritten in terms of the extrema $T_\pm$ of the optimal temperature field, i.e., the zeros of the quartic polynomial $J_x^2+K\TJ^2-\frac{\TJ^4}{A^2}{\bm{J_\perp}^2}$, resulting in
\begin{equation}
\frac{d\TJ}{dx}=2s |J_x| \left[(1-\eta_+\TJ^2)(1+\eta_-\TJ^2)\right]^{1/2}, \label{Tderapp2}
\end{equation}
with the definition $\eta_\pm=\pm 1/T_\pm^2$, such that 
\begin{equation}
(\eta_+\eta_-)^{1/2}=\frac{\vert \vJperp\vert}{A\vert J_x\vert}\, ,
\end{equation}
and we consider one of the $\eta$'s constants fixed by boundary conditions. We can integrate Eq. (\ref{Tderapp2}) between two arbitrary spatial points $(x_A,x_B)$ such that the slope sign $s$ is conserved in the interval
\begin{equation}
\int_{\TJ(x_A)}^{\TJ(x_B)}d\TJ \left[(1-\eta_+\TJ^2)(1+\eta_-\TJ^2)\right]^{-1/2}=2s\vert J_x\vert (x_B-x_A)\, .\label{h2}
\end{equation}
It is now natural to transform Eq. (\ref{h2}) into a Jacobi's integral of the first kind $F(\theta;k)$ \cite{gradshteyn07a,byrd71a} by doing the change of variables $\cos\theta=\sqrt{\eta_+}\TJ$, leading to
\begin{equation}
\frac{2sQ_1T_1\sqrt{\eta_+}}{\sqrt{1-k^2}}(x_B-x_A)=F(\theta(x_A);k)-F(\theta(x_B);k)\, ,\label{h3}
\end{equation}
where
\begin{equation}
F(\theta;k)=\int_0^\theta d\bar\theta\frac{1}{(1-k^2\sin^2\bar\theta)^{1/2}}\, ,
\end{equation}
and where we have defined the modulus via $k^2=(1+\eta_+/\eta_-)^{-1}$, and $Q_1=\vert J_x\vert/T_1$. We can invert Eq. (\ref{h3}) by using the relation
\begin{equation}
\cos\theta=\cn(F(\theta;k);k)
\end{equation}
which defines the cosine-amplitude Jacobi function $\cn(u;k)$ of modulus $k$ \cite{gradshteyn07a,byrd71a}, resulting in
\begin{equation}
\tau(x_B)=\frac{1}{T_1\sqrt{\eta_+}}cn\left(-F(\theta(x_A);k)+\frac{2sQ_1T_1\sqrt{\eta_+}}{\sqrt{1-k^2}}(x_B-x_A);k\right)\, ,\label{h4}
\end{equation}
with $\tau(x)=\TJ(x)/T_1$. Since the $\cn(u;k)$ function appearing in Eq. (\ref{h4}) has a positive maximum and a negative minimum, and taking into account that $\tau(x)$ is defined positive, the optimal temperature field presents at most two possible behaviors: (1) a monotonous decreasing profile or (1) a single-maximum profile. We analyze next each case separately.

\subsection{Monotonous profile}

In this case, assuming without loss of generality that $T_0>T_1$, we have that $s=-1$ for all $x\in[0,1]$. Therefore, considering Eq. (\ref{h4}) with $x_A=0$ and $x_B=x$, the optimal temperature field takes the form
\begin{equation}
\tau(x)=\frac{\cn\Big[F_0+(F_1-F_0)x;k\Big]}{\cn(F_1;k)}\, ,\label{h5}
\end{equation}
with $F_{0,1}\ge 0$ two constants. In order to completely compute the temperature field $\TJ(x)$ we need to fix the values of $F_0,F_1$ and $k$ through boundary conditions and the closure equation 
\begin{equation}
A=\int_0^1 dx\, \sigma[\TJ(x)]\, . \label{app-clausure}
\end{equation}
Indeed, taking into account the boundary conditions, both Eq. (\ref{h3}) and the constraint $\tau(0)\equiv \tau_0=T_0/T_1$ lead to
\begin{eqnarray}
&\displaystyle Q_1=\frac{\sqrt{1-k^2}(F_1-F_0)}{2~\cn(F_1;k)}  \label{h6}\\
&\displaystyle\tau_0=\frac{\cn(F_0;k)}{\cn(F_1;k)}\label{app4},
\end{eqnarray}
respectively, and from (\ref{app-clausure}) we obtain
\begin{equation}
Q_\perp=\frac{E_1-E_0-(1-k^2)(F_1-F_0)}{2k~\cn(F_1;k)}\, ,\label{h8}
\end{equation}
with $Q_\perp=\vert \vJperp\vert/T_1$, and where
\begin{equation}
E_{0,1}=E(\am(F_{0,1};k);k)\, ,
\end{equation}
where $\am(u;k)$ is the amplitude Jacobi function and $E(\theta;k)$ the Jacobi integral of the second kind \cite{gradshteyn07a,byrd71a}. Remarkably, as a consequence of assuming $\tau_0\geq 1$, we find that $F_0\in[0,{\bf K}(k)]$, $F_1\in [\cn^{-1}(\tau_0^{-1};k),{\bf K}(k)]$ with $F_1\geq F_0$ and ${\bf K}$ the Jacobi complete elliptic integral of the first kind (using the notation used by Gradshteyn and Ryzhik \cite{gradshteyn07a}).

\subsection{Single-maximum profile}

In this case $s=+1$ for $x\in[0,x^*]$ while $s=-1$ for $x\in[x^*,1]$, where $x^*$ is the maximum location where $d\tau(x)/dx\vert_{x=x^*}=0$. This maximum position can be obtained from Eq.(\ref{h4}) by taking $x_A=0$, $x_B=x^*$ and forcing the argument of $\cn(u;k)$ to be equal to zero, arriving at
\begin{equation}
x^*=\frac{F_0\sqrt{1-k^2}}{2Q_1T_1\sqrt{\eta_+}}\, .\label{h9}
\end{equation}
In this way, the optimal temperature profile can be determined by considering Eq. (\ref{h4}) for both $x>x^*$ and $x<x^*$, resulting in
\begin{equation}
\tau(x)=\frac{\cn\Big[-F_0+(F_1+F_0)x;k\Big]}{\cn(F_1;k)}\, , \label{appp1}
\end{equation}
where the values of $F_0,F_1\ge 0$ and $k$ are fixed again by the boundary conditions and the closure equation (\ref{app-clausure}), leading again to Eq. (\ref{app4}) and
\begin{eqnarray}
Q_1&=&\frac{\sqrt{1-k^2}(F_1+F_0)}{2~\cn(F_1;k)}\label{appp2}\\
Q_\perp&=&\frac{E_1+E_0-(1-k^2)(F_1+F_0)}{2k~\cn(F_1;k)}\, .\label{appp3}
\end{eqnarray}
Interestingly, Eqs.(\ref{appp1}), (\ref{appp2}), and (\ref{appp3}) corresponding to the single-maximum case map onto Eqs. (\ref{h5}), (\ref{h6}) and (\ref{h8}) corresponding to the monotonous behavior by changing $F_0\rightarrow -F_0$; this allows us to write both solutions in an unified way. In particular, from now on, given $Q_1$, $Q_\perp$ and $\tau_0$ fixed by boundary conditions and Eq. (\ref{app-clausure}), the values of $F_0$, $F_1$ and $k$ are determined from equations (\ref{app4}), (\ref{appp2}), and (\ref{appp3}) with $F_0\in[-{\bf K}(k),{\bf K}(k)]$, $F_1\in [\cn^{-1}(\tau_0^{-1};k),{\bf K}(k)]$ and $k\in[0,1]$ which includes both possibilities: $F_0<0$ for the monotonous case and $F_0>0$ for the single-maximum case.

\subsection{Convexity behavior}

Furthermore, once the solution (\ref{appp1}) has been determined, we can now characterize the convexity behavior of the optimal temperature field as a function of the parameters $(\tau_0,Q_1,Q_\perp)$. Indeed, since the second derivative of $\cn(u;k)$ takes the form $d^2\cn(u;k)/du^2=-\cn(u;k)(1-2k^2+2k^2~\cn(u;k))$, there are no inflection points for $k^2<1/2$ and the profile is always concave in this regime. For $k^2>1/2$ we observe that, for a fixed value of $\tau_0$, both $(Q_1,Q_\perp)$ are parametrized by $(F_0,k)$, leading to the following regions with different number of inflection points. First, for {\bf $\tau_0\leq\sqrt{2}$} we have
\begin{itemize}
\item If $F_0>F_0^{(1)}(k)$ and $k^2\geq1/2$, with $F_0^{(1)}(k)=\cn^{-1}(B(k);k)$ and $B(k)=((2k^2-1)/(2k^2))^{1/2}$, the optimal temperature profile presents two inflection points located at
\begin{equation}
x_{1,2}=(F_0\pm F_0^{(1)}(k))/(F_0+\cn^{-1}(1/\tau_0;k))\, .
\end{equation}
\item If $F_0^{(2)}\leq \vert F_0\vert\leq F_0^{(1)}$ and $k^2\geq1/2$, with $ F_0^{(2)}(k)=\cn^{-1}(\tau_0 B(k);k)$, the optimal profile presents only one inflection point located at $x_1$.
\end{itemize}
 
On the other hand, for {\bf $\tau_0\geq\sqrt{2}$}
\begin{itemize}
\item If $F_0>F_0^{(1)}(k)$ and $k^2\geq1/2$, we find that the profiles present two inflection points at $x_{1}$ and $x_2$.
\item If $F_0^{(2)}\leq \vert F_0\vert\leq F_0^{(1)}$ and $1/2\leq k^2\leq \tau_0^2/(2(\tau_0^2-1))$, the  optimal temperature profile has only one inflection point located at $x_1$. 
\item If $0\leq\vert F_0\vert\leq F_0^{(1)}$ and  $1\geq k^2\geq \tau_0^2/(2(\tau_0^2-1))$, the optimal profile presents again only one inflection point located at $x_1$.
\end{itemize}
 
Finally, outside these regions, no inflection points appear for any value of the parameters.

\section{Limiting Cases of LDF}\label{app3}

In this appendix we study the behavior of the heat current large deviation function both near the steady-state current and in the far tails corresponding to rare current fluctuations. For that, we start from the exact expression for the current LDF obtained in the main text, namely
\begin{equation}
G(\vJJ)=\frac{J_x}{2}\left(\frac{1}{T_0}-\frac{1}{T_1}\right)+\frac{1}{8}(F_0+F_1)^2+\frac{1}{4}(F_0+F_1)\left(\frac{\sn(F_1;k)~\dn(F_1;k)}{\cn(F_1;k)}+\frac{\sn(F_0;k)~\dn(F_0;k)}{\cn(F_0;k)}-E_0-E_1\right)\, ,\label{gtot-app}
\end{equation}
with the different parameters (which depend on $\vJJ$) defined above.

\subsection{Fluctuations around the stationary state}

First, let us introduce the reduced current vector ${\bm{{Q}}}=({Q}_1,{Q}_\perp)$, with the definitions ${Q}_1\equiv |J_x| /T_1$, ${Q}_\perp\equiv |{J_\perp}|/T_1$. For a fixed $k$, it is easy to show that the convergence to the stationary value ${\bm{Q_{st}}}=((\tau_0-1)/2,0)$ takes place when $F_0\rightarrow-{\bf K}(k)$ (see Sec. \ref{sect4}). As a consequence, the behavior of the current LDF near the stationary state can be analyze by fixing $F_0=-{\bf K}(k)+\epsilon$ for small values of $\epsilon$ and any $k$-value. Expanding Eqs. (\ref{p2}) and (\ref{p3}) of the main text around $\epsilon=0$ we realize that they have the structure $\tilde Q_\alpha=Q_\alpha-Q_\alpha^{st}=a_0 \epsilon^2(1+a_1\epsilon^2+\ldots)$, with $\alpha=1,\perp$. It hence seems reasonable to parametrize $\tilde Q_1=R\sin\theta$ and $\tilde Q_{\perp}=R\cos\theta$ and rewrite the expressions as functions of $R$ and $\theta$. Afterwards, we expand $\tilde Q_1/\tilde Q_\perp=\tan\theta$ in terms of $\epsilon$ and look for the $k$-expansion on $\epsilon$ compatible with such expansion and  whose coefficients are functions of $\tan\theta$. Then we substitute the $k$-expansion on the $\tilde Q_1$ expansion and invert the series to find $\epsilon^2$ and $k^2$ as a series expansion on $R$. In particular, we find
\begin{eqnarray}
k^2&=&\frac{1}{2}(1-\sin\theta)+\frac{9(1+\tau_0+\tau_0^2+\tau_0^3+\tau_0^4)}{10(\tau_0-1)(1+\tau_0+\tau_0^2)^2}R\cos^2\theta+O(R^2)\nonumber\\
\epsilon^2&=&\frac{12\tau_0^2}{\tau_0^3-1}R-\frac{12\tau_0^2(27+27\tau_0+7\tau_0^2+7\tau_0^3+7\tau_0^4)}{5(\tau_0-1)^2(1+\tau_0+\tau_0^2)^3}R^2\sin\theta+O(R^3)\, ,\label{exp}
\end{eqnarray}
It is important to note that the expansions are well defined whenever $R/(\tau_0-1)<1$, implying the equilibrium limit ($\tau_0\rightarrow 1$) is singular and cannot be studied by an analytical continuation of the nonequilibrium steady state using Eqs. (\ref{exp}). Substituting these expansion on the expression for the current LDF we find
\begin{eqnarray}
G({\bm{\tilde Q}})&=&\frac{3(\tilde Q_1^2+\tilde Q_\perp^2)}{2(1+\tau_0+\tau_0^2)}\biggl[1-\frac{2(\tau_0-1)(4+7\tau_0+4\tau_0^2)}{5(1+\tau_0+\tau_0^2)^2}\tilde Q_1\nonumber\\
&-&\frac{9}{175(1+\tau_0+\tau_0^2)^4}\biggl(-5(4+2\tau_0-30\tau_0^2-57\tau_0^3-30\tau_0^4+2\tau_0^5+4\tau_0^6)(\tilde Q_1^2+\tilde Q_\perp^2)\nonumber\\
&+&2(\tau_0-1)^2(16+61\tau_0+91\tau_0^2+61\tau_0^3+16\tau_0^4)(\tilde Q_\perp^2-\tilde Q_1^2)\biggr)
+O({\bm{\tilde Q}}^3)\biggr]\, .\label{appGstat}
\end{eqnarray}
In Fig. \ref{fig4} we show in red the gaussian part of Eq. (\ref{appGstat}), $G_{gauss}({\bm{\tilde Q}})=3(\tilde Q_1^2+\tilde Q_\perp^2)/(2(1+\tau_0+\tau_0^2))$, for $\tau_0=2$, making apparent the non-Gaussian asymmetric structure of the exact current LDF. Moreover, we can use expansions (\ref{exp}) around the stationary state to obtain the leading corrections to the linear steady-state temperature profile for small current fluctuations, obtaining
\begin{equation}
\tau(x)=\tau_0-x(\tau_0-1)+\frac{2(\tau_0-1)}{1+\tau_0+\tau_0^2}x(1-x)(1+2\tau_0-x(\tau_0-1))\tilde Q_1+O({\bm Q}^2)\, .
\end{equation}

\subsection{Far tails of the current LDF}

The fluctuating behavior far from the stationary state ${\bf{J_{st}}}=\left(T_1(\tau_0-1)/2,0\right)$ can be better analyzed in two different limits: (1) $\left(\vert J_x\vert\gg J_x^{st},\vJperp=0\right)$ and (2) $\left(J_x=0,\vert \vJperp\vert\gg 0\right)$. The behavior in the first case, $\left(\vert J_x\vert\gg J_x^{st},\vJperp=0\right)$, can be obtained by plugging $k=0$ into Eq. (\ref{gtot-app}) and expanding the expression around its maximum value using $F_0=\pi/2-\epsilon$, which for $J_x>0$ results in
\begin{equation}
G(J_x,0)\simeq\frac{\pi}{2\epsilon}-\frac{4+\tau_0(\pi^2+4)}{8\tau_0}+\frac{\pi(3+5\tau_0)}{24\tau_0}\epsilon+O(\epsilon^2)\, ,
\end{equation}
with
\begin{equation}
\tilde Q_1=Q_1-Q_1^{st}=\frac{\pi\tau_0}{2\epsilon}\left(1-\frac{2\epsilon}{\pi}+\frac{\epsilon^2}{6}+O(\epsilon^3)\right)\, .
\end{equation}
Inverting this series we obtain
\begin{equation}
\epsilon=\frac{\pi\tau_0}{2\tilde Q_1}-\frac{\pi\tau_0^2}{2\tilde Q_1^2}+\frac{\pi(24+\pi^2)\tau_0^3}{48 \tilde Q_1^3}+O(\tilde Q_1^{-4})\, ,
\end{equation}
which finally leads to
\begin{equation}
G(J_x,0)\simeq\frac{J_x}{T_0}-\frac{\pi^2}{8}+\frac{\pi^2(1+\tau_0)T_1}{16J_x}+O(J_x^{-2})\,, \qquad \text{for } \vert J_x\vert\gg J_x^{st},\vJperp=0.
\end{equation}
In order to find the behavior for $-J_x$ we employ the Gallavotti-Cohen relation $G(\vJJ)-G(-\vJJ)=2\epsilon J_x$:
\begin{equation}
G(-J_x,0)=G(J_x,0)+\frac{J_x}{T_1}-\frac{J_x}{T_0}=\frac{J_x}{T_1}-\frac{\pi^2}{8}+\frac{\pi^2(1+\tau_0)T_1}{16J_x}+O(J_x^{-2}),\quad J_x>0\, .
\end{equation}
Therefore the LDF for large current fluctuations along the gradient direction decays linearly with $J_x$, i.e. much more slowly than the quadratic decay one would expect from a naive Gaussian ansatz. Note also that the asymptotic slopes of $G(J_x,0)$ for positive and negative values of the currents $J_x$ are just the inverse temperatures of the left and right reservoirs, respectively. 

The second limit, $\left(J_x=0,\vert \vJperp\vert\gg 0\right)$, corresponds to $k^2\rightarrow 1$ and $F_0\rightarrow\pi/2$, leading to
\begin{equation}
G(0,\vert \vJperp\vert)=\frac{1}{8}\left(F_0+F_1\right)^2\, ,
\end{equation}
with
\begin{equation}
Q_1=0\quad,\quad Q_\perp=\cosh F_1(\tanh F_0+\tanh F_1)/2\quad,\quad \tau=\cosh F_1/\cosh F_0\, .
\end{equation}
Expanding these expressions for large values of $\vert {\vJperp}\vert$, we obtain the following asymptotic form for the LDF
\begin{equation}
G(0,\vert \vJperp\vert)=\frac{1}{8}\ln\left(\frac{4\vert {\vJperp}\vert^2T_1^3}{T_0}\right)\left[\ln\left(\frac{4\vert {\vJperp}\vert^2T_1^3}{T_0}\right)+\frac{1+\tau_0^2}{T_1^2 \vert {\vJperp}\vert^2}+O(\vert {\vJperp}\vert^{-4})\right].
\end{equation}
Remarkably, $G(0,\vert \vJperp\vert)$ exhibits a logarithmic behavior for large current fluctuations orthogonal to the gradient direction. This intricate structure points out to the surprisingly complex analytic behavior of the large deviation function function.

\end{document}